\documentclass[article,twocolumn,showpacs]{revtex4}
\usepackage{
  amsmath,
  amssymb,
  amsthm,
  bm,
  dcolumn,
  fancyvrb,
  framed,
  graphicx,
  hyperref,
  mathtools,
  morefloats,
  textcomp,
  url,
}
\usepackage[normalem]{ulem}
\usepackage[caption=false]{subfig}

\newcommand{\sm}[1]{\mbox{\tiny #1}}

\begin{document}
\title{Dispersion-induced dynamics of coupled modes in a semiconductor laser with saturable absorption}

\author{Finbarr O'Callaghan}
\author{Simon Osborne}
\author{Stephen O'Brien}
\affiliation{Tyndall National Institute, Lee Maltings, University College Cork, Cork, Ireland}

\begin{abstract}
  We present an experimental and theoretical study of modal nonlinear dynamics
  in a specially designed dual-mode semiconductor Fabry-Perot laser with a
  saturable absorber. At zero bias applied to the absorber section, we have
  found that with increasing device current, single mode self-pulsations evolve
  into a complex dynamical state where the total intensity experiences regular
  bursts of pulsations on a constant background. Spectrally resolved
  measurements reveal that in this state the individual modes of the device can
  follow highly symmetric but oppositely directed spiralling orbits. Using a
  generalization of the rate equation description of a semiconductor laser with
  saturable absorption to the multimode case, we show that these orbits appear
  as a consequence of the interplay between the material dispersion in the gain
  and absorber sections of the laser. Our results provide insights into the
  factors that determine the stability of multimode states in these systems, and
  they can inform the development of semiconductor mode-locked lasers with
  tailored spectra.
\end{abstract}


\maketitle

\section{Introduction}
Semiconductor lasers with a saturable absorber can generate short and high-power
optical pulses by the mechanisms of self-pulsation and mode-locking
\cite{vasilev_book, miyajima_09, haus_00, avrutin_00}. These modes of operation
are typically associated with different timescales determined by the relaxation
oscillation frequency [GHz] and the round trip time in the cavity [10-100s GHz].
As self-pulsations (SPs) are often a significant source of instability in
mode-locked lasers, a thorough understanding of mechanisms leading to their
appearance is desirable \cite{rachinskii_06, kolokolnikov_06}.

The model of a laser with a saturable absorber (LSA model) considers the dynamics of the
total field on time scales long compared to the round trip time in the cavity.
While it cannot therefore describe phenomena such as mode-locking, this model
has provided valuable insights into the origins of SPs and the factors that lead
to the appearance of bistability in devices with saturable absorbers
\cite{ueno_85, dubbeldam_99, tronciu_03, dorsaz_11}.

Quantitative dynamical models of semiconductor lasers with saturable absorbers
include travelling-wave methods that consider the spatio-temporal dynamics of
the slowly-varying electric fields \cite{mulet_06, javaloyes_10}. A lumped
element time domain model has also been developed that eliminates the spatial
dependence in favour of a delay-differential equation for the field variable
\cite{vladimirov_04}. These models are efficient tools for understanding the
physical origins of complex spectral and pulse shaping mechanisms in mode-locked
semiconductor lasers \cite{stolarz_11, vladimirov_05, otto_12}.

For certain applications of these devices however, we may be interested in
quantities such as the frequency and phase-noise properties of the individual
locked modes rather than the pulse train generated by the device. Examples
include stable terahertz frequency generation and tailored comb-line emission
demonstrated recently by our group \cite{obrien_10, bitauld_10, bitauld_11a}.
These Fabry-Perot (FP) lasers included a spectral filter to limit the number of
active modes, and here it may be appropriate to formulate the problem of
describing the dynamics in the frequency domain. In such a model, each
longitudinal mode of the cavity is considered as an independent dynamical
variable \cite{lau_90}. The round-trip time in the cavity then determines the
mode spacing, and by including phase-sensitive modal interactions in the model,
one can describe mode-locked states as mutually injection locked steady-states
of the system \cite{avrutin_03}.

In this paper we consider a device that supports two longitudinal modes with a
large frequency spacing. In this case, a frequency domain description based on
an extension of the LSA model represents a natural starting point. A transition
to mode-locking is not possible in this device. Instead, we have found familiar
single-mode SP dynamics, but also interesting examples of coupled dual-mode
dynamics. Here we describe a transition to a multimode state where the total
intensity experiences bursts of fast pulsations. We show that in this state the
individual modes follow oppositely directed spiralling orbits that are related
to the underlying SP dynamics of the system. Our modeling approach is valid for 
small values of the gain and
loss per cavity round trip. Because these approximations are
not expected to hold in a semiconductor laser, our results are
necessarily qualitative. However, we highlight an interesting
example of a multimode instability that can arise in two-section
devices with large dispersion, and our results can guide the
future development of optimized mode-locked devices with
tailored spectra.

This paper is organised as follows. In section \ref{sec:experiment} we introduce
our device and experimental setup. We present optical and mode resolved power
spectra as well as a series of characteristic intensity time traces illustrating
a progression to a region of complex dual-mode dynamics. In section
\ref{sec:modelling} we describe our proposed model equations, which are a
multimode extension of the LSA model that accounts for dispersion of gain and
saturable absorption with wavelength in the system. In section
\ref{sec:analysis} we uncover the bifurcation structure of the system leading to
the measured results. We conclude by discussing the implications of our results
for future work.


\section{Experiment}
\label{sec:experiment}
The device we consider is a multi-quantum well Indium Phosphide based
ridge-waveguide Fabry-Perot laser with one high-reflection (HR) coated mirror.
The total device length is $545$ $\mu$m with a saturable absorber section of
length $30$ $\mu$m adjacent to the HR mirror. The device has a peak gain near
1550 nm, and slotted regions etched in the ridge define a spectral filter, which
is designed to select two primary modes with a spacing of 480 GHz. Further
details on the design of similar devices and their operating characteristics can
be found in \cite{obrien_10}.

\begin{figure}[t]
  \includegraphics{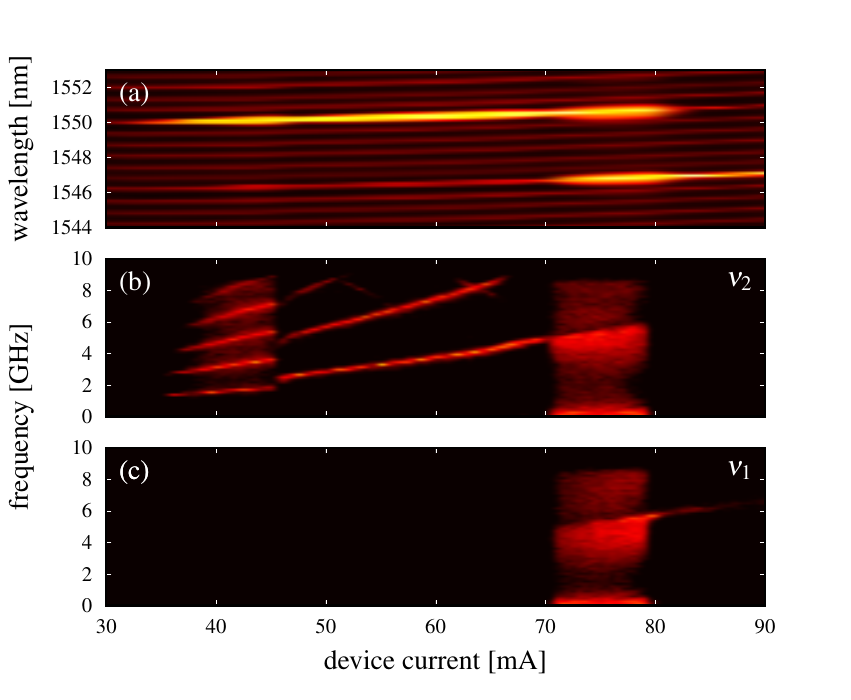}
  \caption{(a) Optical spectrum of the dual-mode two-section
    device as the device current is varied. The bias applied to the absorber
    section is 0 V. (b),(c) Corresponding power spectra of the long
    ($\nu_{2}$) and short ($\nu_{1}$) wavelength primary modes.}
  \label{fig:exptgain}
\end{figure}

Fig. \ref{fig:exptgain} (a) shows the optical spectrum of the laser as the drive
current in the gain section of the device is varied. These spectra were obtained
keeping a constant bias on the short contact of $0$ V and varying the pump
current in the gain section from below lasing threshold at $30$ mA to a value of
$90$ mA. All measurements of this device were carried out at a temperature of
$16.3$\textdegree C.

We label the short and long wavelength primary modes of the device as $\nu_{1}$
and $\nu_{2}$ respectively. From Fig. \ref{fig:exptgain} (a) we see that the
long wavelength mode of the device reaches threshold first at a drive current of
$30$ mA. The two primary lasing modes are located near 1550 nm and 1544 nm and
they have a spacing of six fundamental cavity modes. These primary modes
dominate the spectrum throughout the parameter region of interest. The
corresponding power spectral densities for each of the primary modes are shown
in Fig. \ref{fig:exptgain} (b) and (c). Structure appears in the power spectral
density of $\nu_{2}$ at approximately 35 mA, indicating the onset of dynamical
modulation with a frequency of c. $1.5$ GHz. This transition is also reflected
in a clear spectral broadening visible in the optical spectrum of the mode. We
identify these dynamics as single mode self-pulsations, which appear following a
region of constant output in mode $\nu_{2}$ at threshold. The self-pulsations
are initially sinusoidal and their frequency increases gradually with device
current until a further transition at c. 45 mA. Near this value of the device
current a discontinuity appears in the frequency of the intensity modulation,
which subsequently increases again until a device current of 70 mA, where a
dramatic switch to a region of dual-mode dynamics is observed. In the region of
dual-mode dynamics the power spectra become symmetric with a large range of
frequencies present, including a signature of low-frequency modulation in the
100 MHz range. This region extends over a current range of approximately 10 mA,
with the dynamics switching abruptly to the short wavelength mode $\nu_{1}$ near
80 mA. Following the dual-mode region we observe a single peak in the intensity
power spectrum that gradually diminishes in strength. This indicates that, for
the largest values of the device current shown, we have reached a state of
constant output on the mode at short wavelength.

\begin{figure}[b]
  \subfloat{\includegraphics{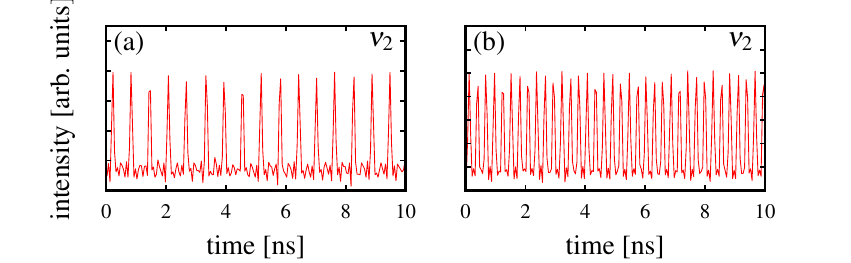}}\\
  \subfloat{\includegraphics{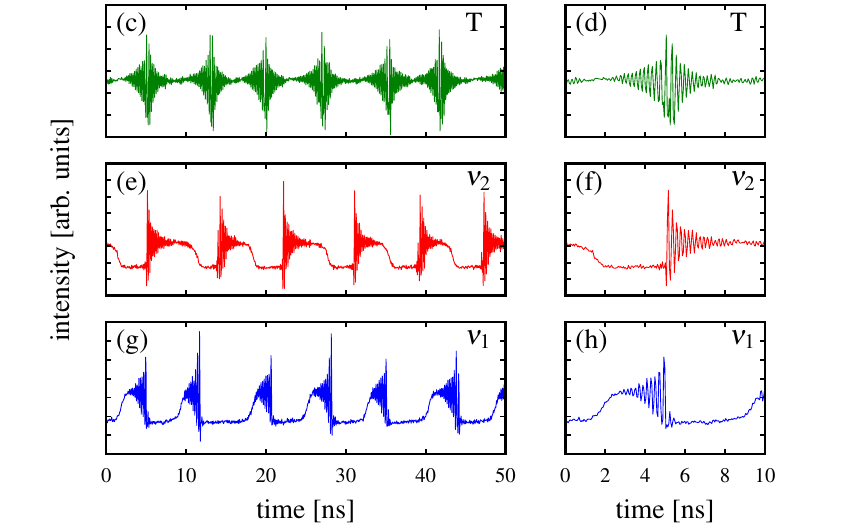}}
  \caption{(a),(b) Measured time traces for the
    long wavelength mode $\nu_{2}$ for a device current in the gain section of
    $44$ and $59$ mA respectively. (c),(d) Time traces of the total
    intensity at a current of $75$ mA in the gain section. (e)-(h) Mode resolved
    time traces at a current of $75$ mA in the gain section. The bias applied to
    the absorber section is $0$ V in all cases.}
  \label{fig:exptttr}
\end{figure}

Representative time traces for the intensity of the long wavelength mode in the
first and second regions of dynamics are shown in Fig. \ref{fig:exptttr} (a) and
(b). The device currents are $44$ and $59$ mA respectively. Fig.
\ref{fig:exptttr} (a) shows characteristic self-pulsation dynamics, where the
intensity reaches small values between pulses and the pulse duration is
significantly less than the interval between pulses. The dynamics in the second
region as shown in Fig. \ref{fig:exptttr} (b) are also strongly modulated but
they are much closer to sinusoidal than in the region of self-pulsations.

Time traces of the total intensity and of the individual modes taken from the
region of dual-mode dynamics are shown in Fig. \ref{fig:exptttr}. (c-h). The
device current in the long contact for these measurements was $75$ mA. One can
see that the total intensity experiences regular bursts of fast pulsations that
are modulated by a much lower frequency envelope. The individual modes in this
dynamical state display a distinctive symmetric saw-tooth structure, where
each mode closely follows a time-reversed trajectory of the other. One can see
that there is a significant antiphase component to these dynamics, as the
intensity of the individual modes reaches values close to zero over a
considerable interval, whereas the total intensity is modulated around a finite
background level.


\section{Modelling of the device response}
\label{sec:modelling}
Modelling the dynamics of our experimental system while treating each mode
individually is complicated by the relatively large number of independent
parameters \cite{javaloyes_10, stolarz_11}. However, we
have successfully modelled the dynamics of dual-mode devices with optical
injection and feedback in past work \cite{osborne_09, brandonisio_12}, and the
LSA model provides a guide for extending these models to the case of a
two-section laser. This model treats the absorber section as an unpumped region
with an unsaturated absorption and carrier recovery time that depend on the
applied voltage.

We do not believe that undertaking a complete theoretical study of the
bifurcation structure of the dual-mode LSA model along the lines of
\cite{dubbeldam_99} would be practical here. Instead, our goal is to understand
the physical roles of the various parameters of the system, and to obtain
numerical estimates for these parameters based on a comparison of simulation
results with the results of our experiment. In physical units the multimode
extension of the LSA model reads
\begin{equation}
  \begin{gathered}
    \dot{S}_m = [(1 - \rho)\tilde{G}_m(N_{g}) + \rho \tilde{A}_m(N_{q}) - \gamma_m] S_m \\
    \dot{N}_{g} = j - \dfrac{N_{g}}{\tau_{s}} - \sum_m \tilde{G}_{m}(N_{g}) S_m \\
    \dot{N}_{q} = \quad -\dfrac{N_{q}}{\tau_{q}} - \sum_m \tilde{A}_m({N}_{q}) S_m
  \end{gathered}
  \label{eq:phys_lsa_2m}
\end{equation}
Here $S_{m}$ is the photon density, and $N_{g}$ and $N_{q}$ are the carrier
densities in the gain and absorber sections respectively. The ratio of the
absorber section length to the total device length is given by $\rho$. The total
field losses of each mode are $\gamma_{m} = \alpha_{\sm{mir}} +
\alpha_{\sm{int}} + \alpha_{\sm{f}}^{m}$, where $\alpha_{\sm{mir}}$ are the
mirror losses and $\alpha_{\sm{int}}$ are the internal losses assumed constant
for all modes. Additional losses, $\alpha_{\sm{f}}^{m}$, due to the action of
the spectral filter are also included. The current density in the gain section
is $j$, while the carrier lifetimes in the gain and absorber sections are
$\tau_{s}$ and $\tau_{q}$ respectively.

\begin{figure}[tb]
  \includegraphics{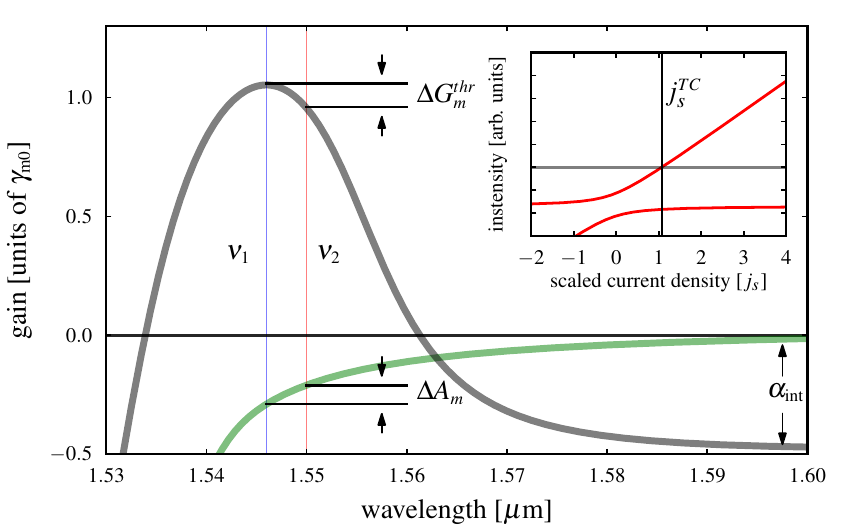}
  \caption{Schematic diagram of the material gain and absorption
    in a typical semiconductor laser. The locations of the two primary modes of
    the device $\nu_1$ and $\nu_2$ are indicated. The model parameters that
    describe the dispersion of the modal gain and absorption are also
    highlighted. Inset: Branches of equilibrium solutions of the single mode LSA
    model for typical parameters considered here. The vertical line is the
    lasing threshold, $j_s^{\mbox{\tiny TC}}$, where the zero field solution
    becomes unstable.}
  \label{fig:gainschematic}
\end{figure}

Typical profiles of the gain and absorption spectra in a semiconductor laser of
the kind we consider are shown schematically in Fig. \ref{fig:gainschematic}.
The negative offset of the gain function at long wavelength gives an estimate of
the background losses, $\alpha_{\sm{int}}$. Here we have indicated the locations
of the two primary modes of the laser. To define the dispersion of gain and
absorption, we first fix a reference carrier density value, $N_{g}^{\sm{thr}}$,
in order to define the dispersion of the gain profile for our model. We take
this reference value to be the threshold carrier density for the device assuming
a transparent absorber section, and define the mode with the largest material
gain at this carrier density as our reference mode, $m_0$. The threshold carrier
density defines a reference value for the modal gain:
$\tilde{G}_{\sm{$m_0$}}(N_{g}^{\sm{thr}}) = (1 - \rho)^{-1}
\gamma_{\sm{$m_0$}}$, and a set of modal differential gain values,
$\tilde{g}_{gm}$. The gain function for each mode can then be linearized around
the reference value of the carrier density so that $\tilde{G}_{m}(N_{g}) =
\tilde{G}_{m}(N_{g}^{\sm{thr}}) + \tilde{g}_{gm} (N_{g} - N_{g}^{\sm{thr}})$.
Dispersion in the modal absorption is included by defining $\tilde{A}_m(N_{q}) =
\tilde{g}_{qm} N_{q} - A_{m}^{0}$, where the differential absorption is
$\tilde{g}_{qm}$, and unsaturated losses for each mode, $A_{m}^{0}$, will be
determined by the applied voltage.

To derive normalized equations, we rescale time in units of the photon decay
rate of a plain FP cavity without spectral filtering: $\gamma =
\alpha_{\sm{mir}} + \alpha_{\sm{int}}$. We define the normalized pump current,
$p = (j - j_{\sm{thr}})/j_{\sm{thr}} \equiv j_s - 1$, where $j_{\sm{thr}} =
N_{g}^{\sm{thr}}/\tau_s$, and we define the normalized carrier densities in each
section of the device: $n_g = (N_g - N_{g}^{\sm{thr}})/N_{g}^{\sm{thr}}$ and
$n_q = (N_0 - N_{q})/N_{g}^{\sm{thr}}$, where $N_0 =
A_{m_0}^{0}/\tilde{g}_{qm_0}$. In normalized units the equations then read
\begin{equation}
  \begin{gathered}
    \dot{I}_m = [(1 - \rho)G_m(n_{g}) + \rho A_m({n}_{q}) - \gamma_m^{'}] I_m \\
    T \dot{n}_{g} = p - n_{g} - \sum_m G_m(n_{g}) I_m \\
    T \dot{n}_{q} = \Delta(q_0 - n_{q}) + \sum_m A_m({n}_{q}) I_m
  \end{gathered}
  \label{eq:lsa_2m}
\end{equation}
where $q_0 = \frac{N_0}{N_{g}^{\sm{thr}}}$, $T = \gamma \tau_s$, and $\Delta =
\tau_s/\tau_q$. In these equations the normalized gain functions are
\begin{equation}
    G_m(n_{g}) = G_m (n_{g}^{\sm{thr}}) + g_{gm} N_{g}^{\sm{thr}} n_{g} \nonumber
\end{equation}
where $G_m (n_{g}^{\sm{thr}}) = \gamma^{-1}\tilde{G}_{m_0}^{\sm{thr}} + \Delta
G_{m}^{\sm{thr}}$, $\gamma_m^{'} = \gamma_m/\gamma$, and $g_{gm} =
\tilde{g}_{gm}/\gamma$. Here $\Delta G_{m}^{\sm{thr}} \equiv \gamma^{-1}
(\tilde{G}_m^{\sm{thr}} - \tilde{G}_{m_0}^{\sm{thr}})$ describes the dispersion
of the reference linear gain profile. The normalized modal absorption functions
are
\begin{equation}
    A({n}_{q}) = - g_{qm} N_{g}^{\sm{thr}} n_{q} + (g_{qm} - g_{q m_0}) N_0 - \gamma^{-1} \Delta A_m^{0} \nonumber
\end{equation}
where $g_{qm} = \tilde{g}_{qm}/\gamma$, and $\Delta A_m^{0} = A_{m}^{0} -
A_{m_0}^{0}$. Here the normalized carrier density in the absorber section
is defined so that the modal absorption is linearized around the saturated value
for the reference mode. Note that the phase space of system \eqref{eq:lsa_2m}
contains two invariant three dimensional sub-manifolds, defined by $I_m = 0$,
for $m = \{1,2\}$. The dynamics on each of these sub-manifolds reduces to the
single-mode LSA system, and for this reason we will refer to these sub-manifolds
as the single-mode manifolds of the system.

Based on previous estimates obtained for similar devices \cite{osborne_09}, we
take the carrier lifetime in the gain section, $\tau_s$ = 1 ns. The mirror
losses of the device are calculated to be $\alpha_m = 13.8 \mbox{ cm}^{-1}$, and
we estimate the internal losses to be $\alpha_{\mbox{\tiny int}}= 9.2
\mbox{ cm}^{-1}$. These losses determine the cavity decay rate for the plain FP
laser to be $\gamma = 2 \times 10^{11} s^{-1}$, and $T \simeq 200$.  In order to
define the losses due to spectral filtering and to fix the value of
$N_{g}^{\sm{thr}}$, we compared our results with threshold data from a plain
two-section FP laser. With a uniform current density over the full device
length, the threshold current of the FP laser was 13.5 mA. With a bias of 0 V
applied to the absorber section, the threshold increased to 17 mA, with the peak
emission at 1557.5 nm. The scale of $N_{g}$ is defined so that the differential
gain at threshold for the single section FP is equal to 1 in normalised units.
We therefore set $N_{g}^{\sm{thr}}$ equal to 2.2 based on estimates of the
increase in the carrier density necessary to reach the defined threshold level
at the wavelength of the reference mode. This estimate was made by taking the
losses due to spectral filtering $\alpha_{f}^{\sm{m}} = 10 \mbox{ cm}^{-1}$, and
using an approximate model for the semiconductor susceptibility \cite{balle_98}.
This model allowed us to account for the large blue-shift of the gain peak from
its position in the FP laser at threshold. We note that the measured increase in
threshold of the dual-mode device, the placement of the spectral filter and the
large separation between the selected modes are all factors that suggest large
unsaturated absorption and enhanced dispersion of the model parameters.

The carrier lifetime in the absorber section can be much shorter than in the
gain section, with a strong dependence on the applied bias \cite{karin_94}. We
fix the carrier lifetime in the absorber section to be 50 ps, so that $\Delta =
20$. However, provided $\Delta$ is not close to one, we have found our results
are not dependent on the precise value of this quantity. In order to complete
the model we must specify the values of the linear gain, unsaturated absorption,
and the differential gain and absorption for each primary mode of the device.
Because of the large size of this parameter space, we begin by considering the
dynamics of two coupled modes with similar parameters. Guided by the known
dispersive properties of the semiconductor susceptibility, and by the observed
behavior of the device, we then make a series of further adjustments to these
parameters until we have obtained satisfactory agreement with measured data.


\section{Bifurcations of a dual-mode semiconductor laser with a saturable absorber}
\label{sec:analysis}

For our numerical simulations, the parameters describing the gain function at
the position of the reference mode, $\nu_1$, are fixed. To begin we assume a
flat gain and absorption curve and we examine the effects of dispersion in the
differential gain and absorption on the dynamics of the coupled system. From
Fig. \ref{fig:exptgain} we see that in our experiment the device begins to lase
with constant intensity output on the long wavelength mode, before entering a
region of SP. At the largest values of the pump current the intensity switches
to short wavelength. With equal linear gain and unsaturated absorption, the mode
with the largest differential gain will reach threshold first. On the other
hand, a larger differential absorption will mean that a mode will saturate its
losses more quickly above threshold and thereby dominate at larger pump values.
To reproduce this behaviour, we set the differential gain and absorption of mode
$\nu_{1}$ in normalised units to be $1.0$ and $1.2 \rho^{-1}$ respectively. The
ratio of these quantities for mode $\nu_{1}$ is then $s_1 \equiv g_{q1}/g_{g1}$
= 22. We set the differential gain and absorption of $\nu_{2}$ to be $1.4$ and
$0.6 \rho^{-1}$ respectively so that $s_2$ = 7.8. The remaining parameters
values we choose to begin are $A_{\sm{[1,2]}}^{0} = 0.2 \gamma$ and $\Delta
G_{m\sm{[1,2]}}^{\sm{thr}} = 0$. Note that the differential gain in normalised
units is likely to be less than unity given the higher current density at
threshold in the dual-mode device. We have decided not to make this correction
in order to make the comparison of the various model parameters more
transparent. We have confirmed that our results are largely independent of the
precise values chosen for $g_{g1}$, provided the other differential quantities
are scaled accordingly.

While the LSA model can predict the appearance of self-pulsations immediately at
threshold \cite{ueno_85, dubbeldam_99}, our chosen parameter values are
consistent with the observation of a narrow region of constant output at
threshold in our experiment. If we consider a single mode system, the zero field
equilibrium solution of these equations is stable until a transcritical
bifurcation at a threshold value of the pump
\begin{displaymath}
p_{\mbox{\tiny TC}} = \frac{\rho A_{m}^{0}}{(1 - \rho) g_{g} \gamma N_{g}^{\sm{thr}}}.
\end{displaymath}
The inset of Fig. \ref{fig:gainschematic} shows the branches of single mode
equilibrium solutions of Eqn \eqref{eq:lsa_2m} taking the model parameters for
mode $\nu_{1}$. Here $\Delta$ exceeds a minimum value given by
\begin{displaymath}
\Delta = s \frac{q_0^{'}}{1 + q_0^{'}}.
\end{displaymath}
where $q_o^{'} = \gamma^{-1} \rho A_m^{\scriptscriptstyle 0}$. This condition leads to constant 
output at threshold, as the upper branch of equilibrium solutions takes physical values
after exchanging stability with the zero field solution. A further bifurcation
to SPs will then occur provided there is sufficient saturable absorption in the
system. We will find that
the stability of the single-mode equilibria plays a fundamental role in
organising the dynamics in our device, and we will therefore present numerical
bifurcation diagrams for each of the single-mode solutions of our model as we
vary the model parameters for each mode.

\begin{figure}[t]
  \includegraphics{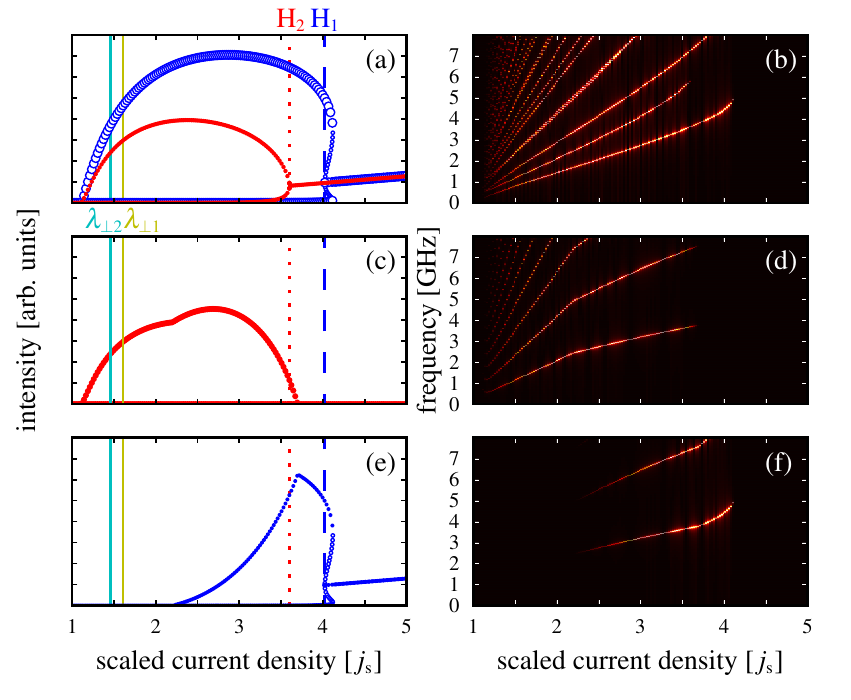}
  \caption{Left panels: Simulated bifurcation diagrams. Right
    panels: Power spectral densities. (a),(b): Single-mode dynamics. (c),(d): 
    Long wavelength mode $\nu_2$. (e),(f): Short wavelength mode
    $\nu_1$.}
  \label{fig:sim_full_inphase}
\end{figure}

Numerical bifurcation diagrams and intensity power spectra obtained with our
first set of parameters are shown in Fig. \ref{fig:sim_full_inphase}. Figs
\ref{fig:sim_full_inphase} (a) and (b) describe the dynamics of both of the
primary modes restricted to their respective single mode manifolds, obtained by
setting the intensity of the inactive mode to zero for the time evolution of Eqn
\eqref{eq:lsa_2m}. In Fig. \ref{fig:sim_full_inphase} (a), as expected, the
single mode dynamics of both modes exhibit threshold behaviour similar to the
observed behaviour of the long wavelength mode in our experiment, with a region
of constant output found after the zero field solution becomes unstable.
Following this region, they enter a region of SPs at the location of the first
Hopf bifurcation, and the SP region is bounded in each case by a second Hopf
bifurcation at larger pump current. In Fig. \ref{fig:sim_full_inphase}, dashed
and dotted lines labelled H$_1$ and H$_2$ indicate the second Hopf bifurcation
points that bound the SP region at larger pump values for each mode.

Mode resolved numerical bifurcation diagrams and power spectra for the two modes
in the full coupled mode system are shown in Fig. \ref{fig:sim_full_inphase}
(c)-(f). Chosen parameters ensure that the long wavelength mode reaches
threshold first, and because the second mode is initially suppressed, mode
$\nu_{2}$ reproduces the dynamics found in the single mode system. However,
before the region of single mode SP ends at the second Hopf bifurcation shown in
Fig. \ref{fig:sim_full_inphase} (a), the dynamics become dual-mode, with the SP
intensity gradually switching across to the shorter wavelength mode $\nu_{1}$ as
the pump is increased further. This dual-mode region comes to an end shortly
after a pump value of $j_{s} = 3$ where the system enters a region of single
mode SP on $\nu_{1}$. The region of SP dynamics finally ends at the subcritical
Hopf bifurcation of $\nu_{1}$, and the dynamics switch to constant output in
mode $\nu_{1}$ for large values of the pump current.

In the power spectrum of Fig. \ref{fig:sim_full_inphase} (d) we see that at the
onset of SPs, they occur with a frequency of c. $500$ MHz, with a linear
increase in each interval of single or dual-mode dynamics thereafter, and
reaching a value of c. 5 GHz at $j_s = 4.25$. We can compare this evolution
with the dependence of the relaxation oscillation frequency, which, neglecting
the effects of saturable absorption, is given by
\begin{equation}
\omega_{\sm{RO}} = \sqrt{\frac{(1-\rho) g_g N_g^{\sm{thr}}p}{T}}.
\end{equation}
At $j_s = 4.25$, the result is approximately 6 GHz, which is a reasonable
estimate of the SP frequency in this model. Note however that the close to
linear increase of the numerical SP frequency contrasts with the square-root
dependence of the above expression.

If we compare the numerical variation of the SP frequency with our experiment,
we see that the measured SP frequency appears with a large value of c. 1.5 GHz,
and that it then remains relatively constant. Our simulations therefore
underestimate the SP frequency at onset, and overestimate its rate of increase.
One can also see that the extent of the SP region that we find numerically is
much larger than the measured value. While the measured variation of the SP
frequency may be partly due to an uncharacteristic behavior of our device, it
should be noted that we cannot expect to obtain quantitative agreement with
experiment using the LSA model. We have confirmed this by comparing the results
of numerical simulations made using the LSA model and the delay-differential
model of \cite{vladimirov_04}. For example, using experimentally calibrated
parameters appropriate to a self-pulsating FP laser, we have found that the LSA
model will in general predict a far larger SP region, with a larger SP frequency
than the delay-differential model. In addition, while we can adjust unknown
parameters such as the absorber recovery time to match numerical and
experimental results in the case of the delay differential model, this is in
general not possible when using the LSA model. This comparison emphasizes the added
importance of accounting for the large changes in gain and
loss that can occur in two-section semiconductor lasers, where
SP dynamics involve strong saturation of the absorption for
typical parameters.

\begin{figure}[b]
  \subfloat{\includegraphics{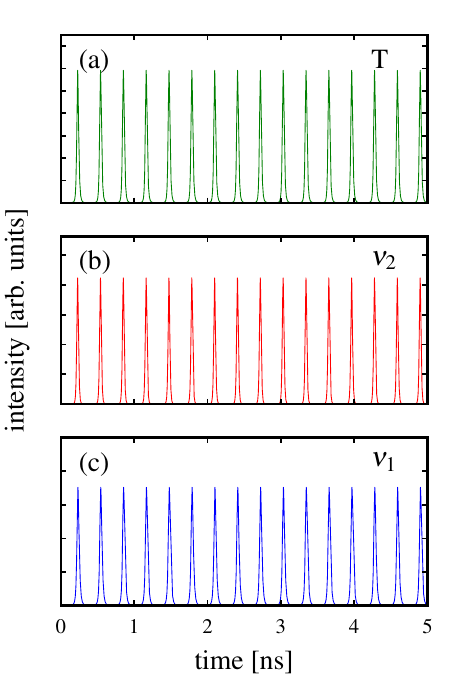}}
  \subfloat{\includegraphics[width=0.49\columnwidth,height=20em]{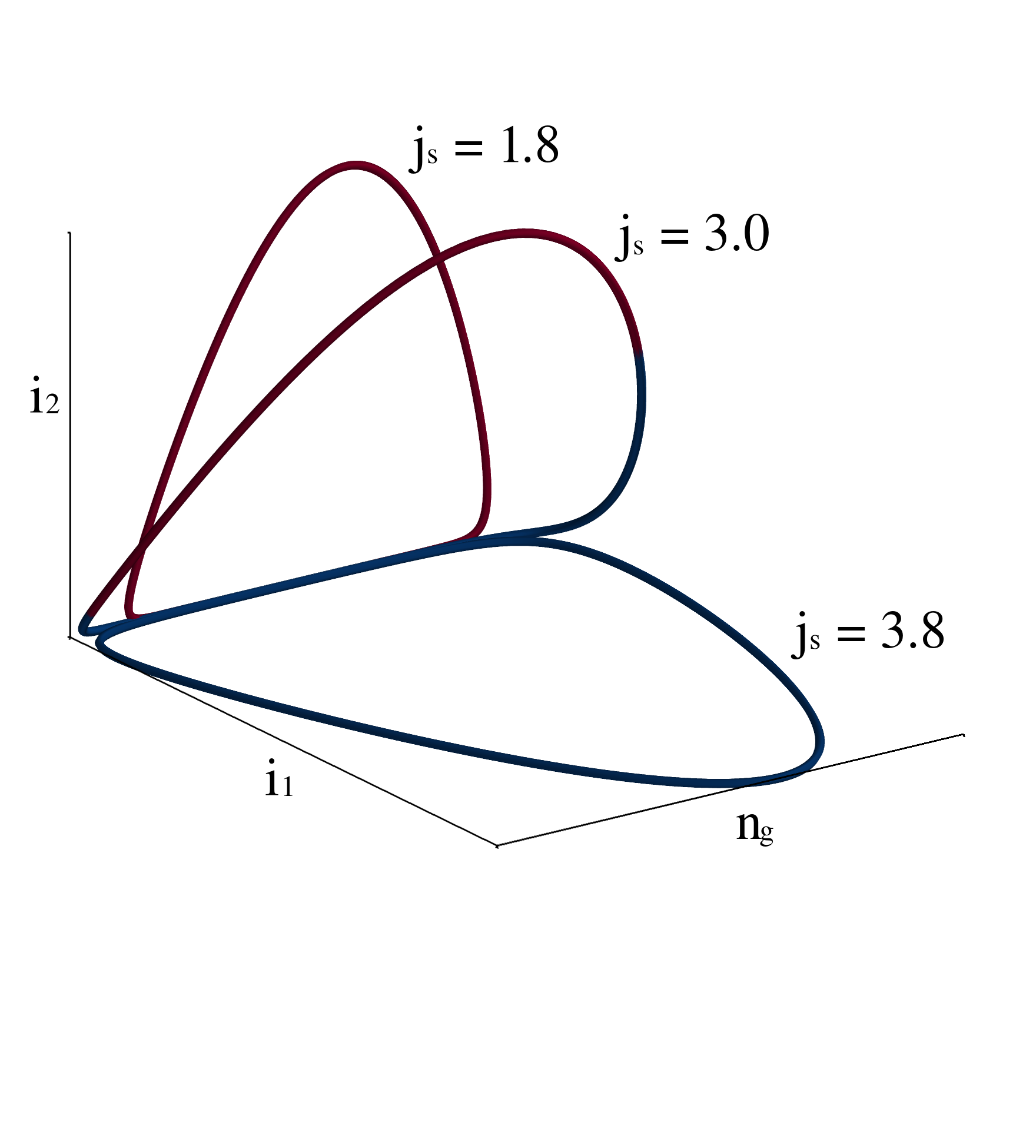}}
  \caption{Left: Simulated time traces of the total intensity
    (upper panel), and of the individual modes (center and lower panels). The
    pump current value is $j_s$ = 3. Right: Phase space diagrams for three
    values of the pump current as shown.}
  \label{fig:sim_inphase}
\end{figure}

Time traces taken from the center of the dual-mode region with $j_{s} = 3$ are
shown in the left panel of Fig. \ref{fig:sim_inphase}, while the right hand
panel of Fig. \ref{fig:sim_inphase} shows a phase space representation of the
dynamics for three pump values, taken before, during and after the transition
from $\nu_{2}$ to $\nu_{1}$. We see that the intensity shifts continuously from
one mode to the other through a region of in-phase SP. As expected, the gradual
transition of the dynamics from $\nu_{2}$ to $\nu_{1}$ that we observe in these
simulations leads to agreement with the experimental measurements of Fig.
\ref{fig:exptgain} near threshold and at large values of the pump.

Valuable insight into what factors can lead to better agreement with experiment
can be gained from a close examination of the time traces presented in Fig.
\ref{fig:exptttr} (c-h). From these figures, we see that for the majority of the
orbit duration the intensity is close to the single mode manifold of one mode or
the other. In Fig. \ref{fig:exptttr} (f) we see the large intensity oscillations
of $\nu_{2}$ decaying toward a state with almost constant output, and the
intensity then quickly switching to a similar state in $\nu_{1}$ from which the
oscillations grow again. The clear observation of this near single mode state
with constant output at the beginning and end of these bursts of pulsations
suggests that the single mode equilibrium states of \eqref{eq:lsa_2m} may be
playing an important role in organising the observed dynamics. In particular,
given the presence of symmetric invariant sub-manifolds in the phase-space, we
should examine the dependence of the transverse stability of the single mode
equilibrium solutions on the model parameters.

\begin{figure}[tb]
  \includegraphics{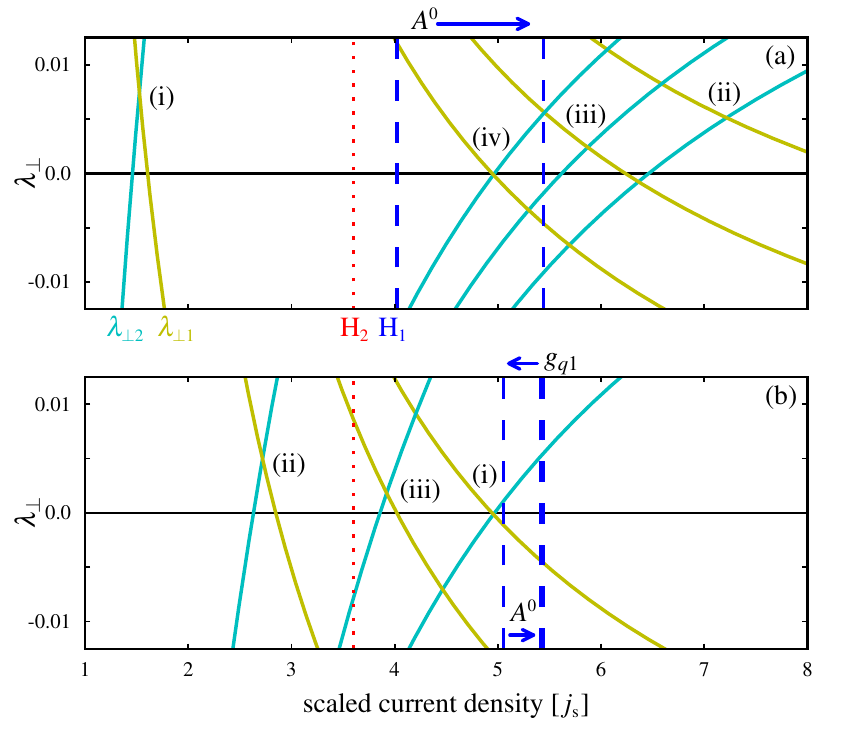}
  \caption{Effect of including dispersion in model parameters on
    the transverse Lyapunov exponent and the location of the second Hopf
    bifurcation of each single-mode equilibrium of the model. Dashed and dotted
    lines indicate Hopf bifurcations of $\nu_{1}$ and $\nu_{2}$ respectively.
    Solid lines are the transverse Lyapunov exponents of each mode as indicated.
    (a) $\text{(i)}$ Parameters as in Fig. \ref{fig:sim_full_inphase},
    $\text{(ii)-(iv)} A_{\sm{1}}^{0} = 0.35 \gamma$ and $\Delta
    G_{m2}^{\sm{thr}} = [\text{(ii)} 0, \text{(iii)} -0.25, \text{(iv)}
    -0.5]\gamma^{-1}$. (b) $\text{(i)}$ Parameters as in (iv) of (a),
    $\text{(ii), (iii) } g_{q1} = 1.5$, and $A_{\sm{1}}^{0} = \text{(ii)} 0.35
    \gamma, \:\text{(iii)} 0.4 \gamma$.}
  \label{fig:lyap_prog1}
\end{figure}

The transverse stability of the single-mode equilibrium solution for mode
$\nu_{i}$, $I_{i}^{0}$, is determined by the sign of its transverse Lyapunov
exponent
\begin{equation}
  \lambda_{\perp i} = (1-\rho)G_{j}(I_{i}^{0}) + \rho A_{j}(I_{i}^{0}) - \gamma_{j}
  \label{eq:lyap}
\end{equation}
where $\{i,j\} = \{1,2\}$ and mode $\nu_{i}$ is transversally stable for
negative values of $\lambda_{\perp i}$. For illustration purposes, we have
plotted and labelled solid lines in Fig. \ref{fig:sim_full_inphase} that
indicate where the changes in transverse stability occur for the chosen
parameter values. Note that in this case $\nu_{2}$ becomes transversally
unstable while $\nu_{1}$ becomes transversally stable with increasing $j_{s}$.
The impact of dispersion of linear gain and unsaturated absorption on the
dynamics of our model can be illustrated by a plot of the locations of the Hopf
bifurcations and changes in transverse stability of the single mode equilibrium
solutions as the relevant parameters are varied. In Fig.
\ref{fig:gainschematic}, the physical dispersion of the unsaturated absorption
suggests a larger value of $A^{0}$ for $\nu_{1}$, and stability changes for both
modes obtained with $A_{\sm{1}}^{0}$ increased to a value of $0.35 \gamma$, and
$\Delta G_{m2}^{\sm{thr}}$ ranging from $0$ to $-0.5 \gamma^{-1}$ are shown in
Fig. \ref{fig:lyap_prog1} (a). In these figures vertical dashed and dotted lines
indicate the locations of the second Hopf bifurcation for $\nu_{1}$ and
$\nu_{2}$ respectively. Curved solid lines plot the value of the transverse
Lyapunov exponent for each mode as indicated, with sign changes of these
quantities indicating changes in transverse stability. One can see the impact
that a change of only $0.5 \mbox{ cm}^{-1}$ to the linear gain profile has on
the transverse stability properties of these modes. The cumulative net effect of
these changes is that there is now a much larger separation between the second
Hopf bifurcation points of both modes. In addition, the sign changes of the
transverse Lyapunov exponents for each mode now occur between the pair of Hopf
bifurcations.

\begin{figure}[t]
  \includegraphics{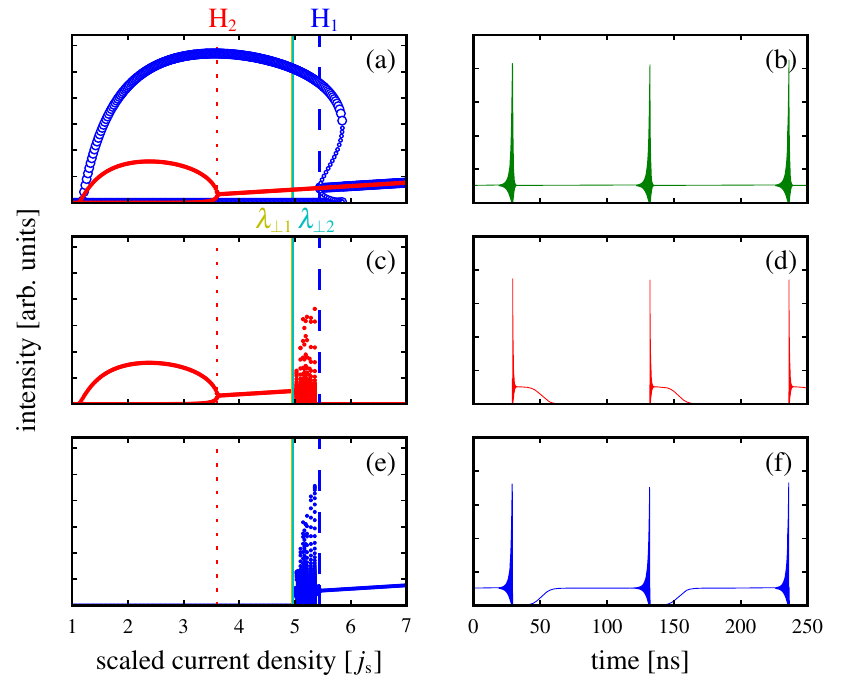}
  \caption{Left panels: Simulated bifurcation diagrams. Right
    panels: Intensity time traces. (a) Single-mode dynamics. (b) Total
    intensity. (c),(d) Long wavelength mode $\nu_2$. (e),(f)
    Short wavelength mode $\nu_1$.}
  \label{fig:sim_full_cw-qs}
\end{figure}

Numerical bifurcation diagrams obtained for parameter set (iv) of Fig.
\ref{fig:lyap_prog1} (a) are plotted in in the left hand panels of Fig.
\ref{fig:sim_full_cw-qs}. Here, $A_{\sm{1}^{0}} = 0.35 \gamma$, $\Delta
G_{m\sm{2} }^{\sm{thr}} = -0.5 \gamma^{-1}$, and all other parameters are
unchanged from Fig. \ref{fig:sim_full_inphase}. The increased unsaturated losses
mean that mode $\nu_1$ is suppressed for longer and, in contrast to the results
of Fig. \ref{fig:sim_full_inphase}, mode $\nu_{2}$ now completes a region of
single-mode SP bounded by two single mode Hopf bifurcations. Following the SP
region, the system enters a region of single-mode constant output. However, as
we increase the pump current still further we find the dynamics are dramatically
``blown-out'' from the single mode manifold and we enter a region of coupled
dynamics in both modes. The location of the blow-out is determined by the loss
of transverse stability of $\nu_{2}$, and this point is located shortly after
$\nu_{1}$ has become transversally stable. The region of dual-mode dynamics is
bounded at larger values of the pump current by a subcritical Hopf bifurcation
of mode $\nu_{1}$.

In order to compare the measurements of Fig. \ref{fig:exptttr} in the dual-mode
region with the simulation results of Fig. \ref{fig:sim_full_cw-qs}, we have
plotted mode-resolved and total intensity time traces with $j_{s}$ = 5.1 in the
right-hand panels of Fig. \ref{fig:sim_full_cw-qs}. The similar nature of the
two dynamical states is clear, with the numerical results reproducing the
observed bursts of fast pulsations in the total intensity and also the switching
sequence with the intensity rising on the short wavelength mode before switching
and falling on the long wavelength mode. However, the frequency of the bursts
that we find numerically is much too low at around 10 MHz. In addition, the
numerical bifurcation sequence in not in full agreement with our measurements.
We find a large region of constant output between the second Hopf bifurcation of
$\nu_{2}$ and the onset of coupled mode dynamics and we also do not see any
evidence of a discontinuity in the frequency of the intensity modulation at
intermediate values of the pump.

\begin{figure}[b]
  \includegraphics{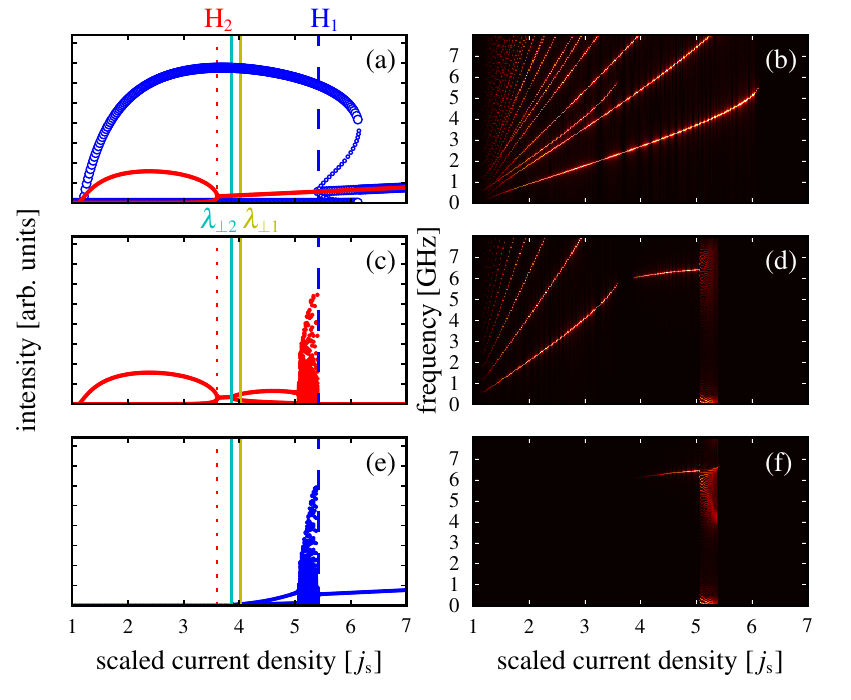}
  \caption{Left panels: simulated bifurcation diagrams. Right
    panels: Power spectral densities. (a),(b): Single-mode dynamics. (c),(d): 
    Long wavelength mode $\nu_2$. (e),(f): Short wavelength mode
    $\nu_1$.}
  \label{fig:sim_full_qs-qs}
\end{figure}

By further adjustment of parameters we can shift the point where the transverse
stability of the equilibrium state of $\nu_1$ changes at smaller values of the
pump. This will result in a narrowing of the region of constant output in
agreement with experiment. Fig. \ref{fig:lyap_prog1} (b) illustrates the effect
of an increase in $g_{q1}$ combined with a further increase of $A_{\sm{1}}^{0}$.
The net effect of these adjustments is that the positions of the Hopf
bifurcations remain largely the same, but the changes of transverse stability
happen much closer in pump current to the second Hopf bifurcation of $\nu_{2}$.
Numerical bifurcation diagrams and intensity power spectra with $g_{q1} = 1.5$
and $A_{\sm{1}}^{0} = 0.4 $ are shown in Fig. \ref{fig:sim_full_qs-qs}. The
bifurcation sequence we observe until the change in transverse stability of the
single mode equilibrium of $\nu_2$ is the same as in Fig.
\ref{fig:sim_full_cw-qs}. However, instead of a dramatic transition to a region
of complex coupled dynamics, in this case we find a very narrow region where a
dual-mode equilibrium state of the system is stable. This state appears here for
the first time in our simulations, and it appears because the order of the
changes in transverse stability of the single-mode equilbiria has been reversed
compared to the previous example. We find that the dual-mode equilibrium state
quickly evolves into a dual-mode limit cycle at a Hopf bifurcation point. This
dual-mode limit cycle is unusual in that the amplitude of $\nu_1$ over the cycle
is very weak to begin. With a further increase in the pump current, the dual
mode limit cycle loses stability, and we observe a dramatic transition to a
region of complex coupled dynamics in both modes. As in the previous example,
the region of complex dynamics is bounded at a large pump current by a
subcritical Hopf bifurcation in $\nu_{1}$.

\begin{figure}[t]
  \subfloat{\includegraphics{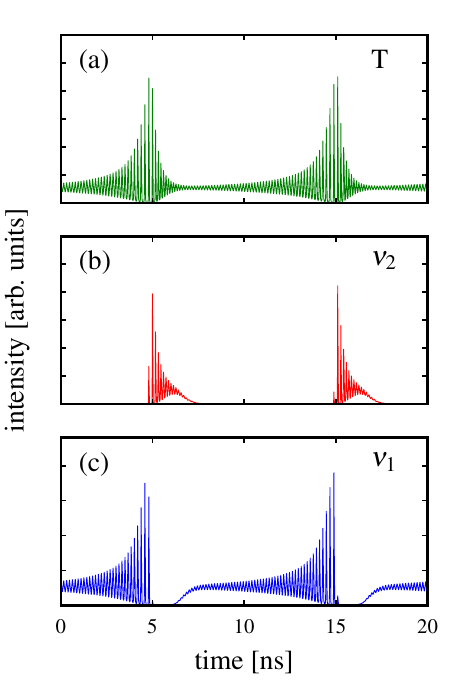}}
  \subfloat{\includegraphics[width=0.49\columnwidth,height=20em]{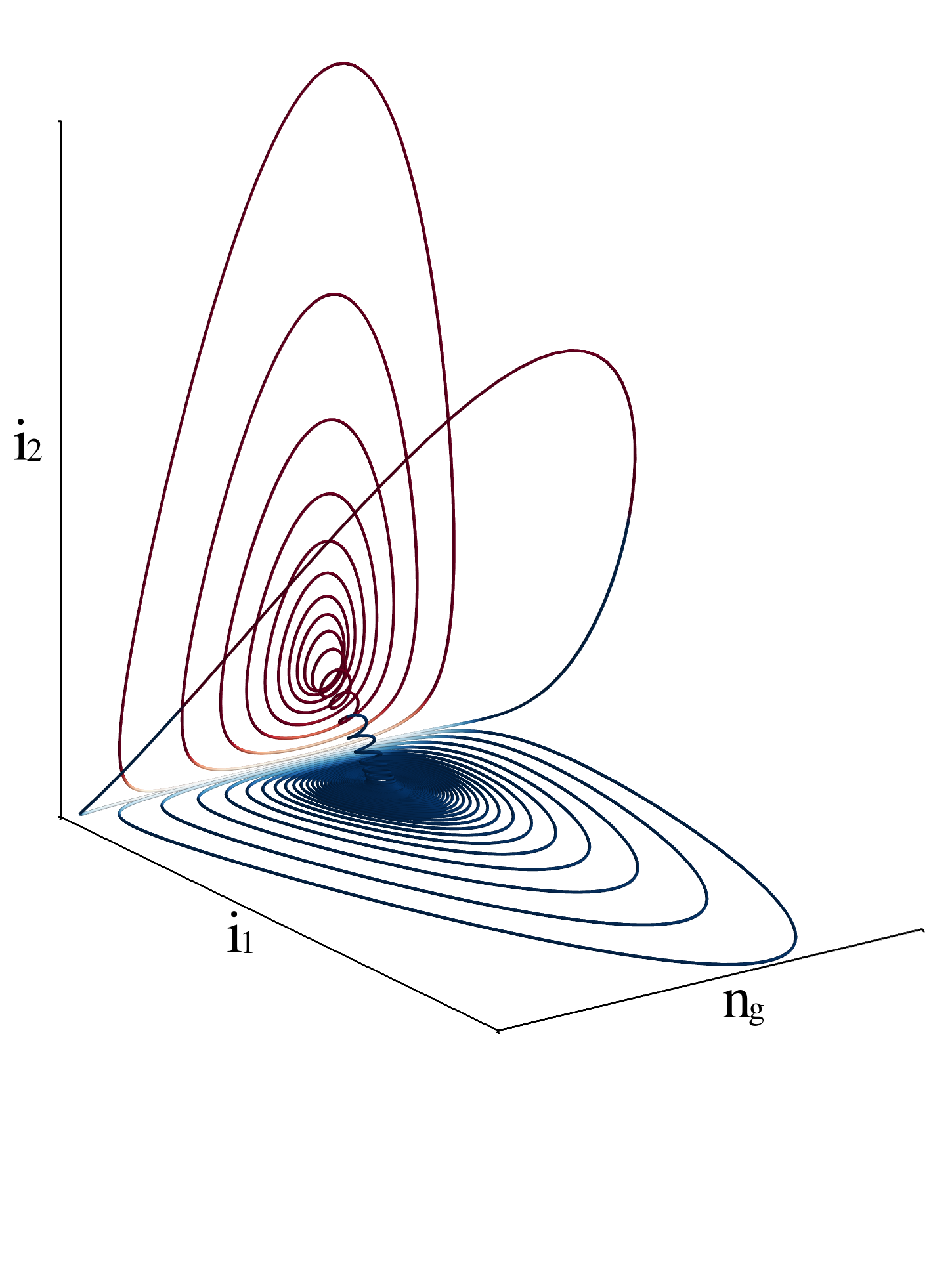}}
  \caption{Left: Simulated intensity time traces. (a) Total
    intensity. (b) Long wavelength mode $\nu_2$. (c) Short wavelength mode
    $\nu_1$. Right: Phase space diagram}
  \label{fig:sim_qs-qs}
\end{figure}

A plot of the mode-resolved and total intensity time traces taken from the
region of complex coupled dynamics in Fig. \ref{fig:sim_full_qs-qs} with $j_{s}$
= 5.2 is shown in the left-hand panels of Fig. \ref{fig:sim_qs-qs}. When
compared to our experimental results, there is a greater degree of asymmetry
between the dynamics of the two modes in this example. Unlike the previous
example however, the frequency of the bursts of fast pulsations of the total
intensity is now accurately matched to our experimental results. We note also
that the observed bifurcation sequence provides an explanation for the dynamics
we found at intermediate values of the pump current in our experiment. We can
now see that the discontinuity in the frequency of the intensity modulation of
$\nu_2$ was due to the momentary appearance of a stable dual-mode steady state
in the system, leading to the absence of any structure in the intensity power
spectrum over this interval. In addition, we must reinterpret the region after
the discontinuity as a dual-mode state with strong intensity modulation, but
where the amplitude of the component in $\nu_1$ is very weak.

The time-traces of Fig. \ref{fig:sim_qs-qs} are depicted in a phase-space
representation on the right of Fig. \ref{fig:sim_qs-qs}. If we consider
trajectories close to the single-mode equilibrium state of $\nu_1$, these
trajectories are attracted towards the single mode manifold as the single-mode
equilibrium state has a negative transverse Lyapunov exponent. Because this
state is unstable in the single-mode manifold, as the trajectory approaches the
manifold it is repelled into a spiralling orbit towards the SP limit cycle,
which is stable within the $\nu_1$ manifold. This is the origin of the fast
pulsations that grow from the quasi-single mode steady state of $\nu_1$. Once
the trajectory approaches the SP limit cycle, it feels the transverse
instability of this limit cycle and is ultimately ejected from the region near
the single-mode manifold, undergoing a large amplitude excursion where both
fields have large intensity. This large excursion leads the trajectory to enter
the slow region near zero intensity, where it is drawn towards the single mode
manifold of $\nu_2$, where the single mode equilibrium state is stable within
the manifold. This leads to the cycle of fast pulsations that decay towards the
single mode equilibrium state of $\nu_2$. Finally, as the trajectory approaches
the equilibrium, it is repelled on account of the positive transverse Lyapunov
exponent of this state. This repulsion drives the trajectory towards the
transversally attracting equilibrium state of $\nu_1$ and the cycle begins
again. The switch in intensity from $\nu_2$ to $\nu_1$ is along a corkscrew type
trajectory that is wound around the line connecting the equilibrium states in
the two single mode manifolds. This winding may be a signature of the dual-mode
limit cycle present before the region of complex dynamics.

\begin{figure}[t]
  \includegraphics{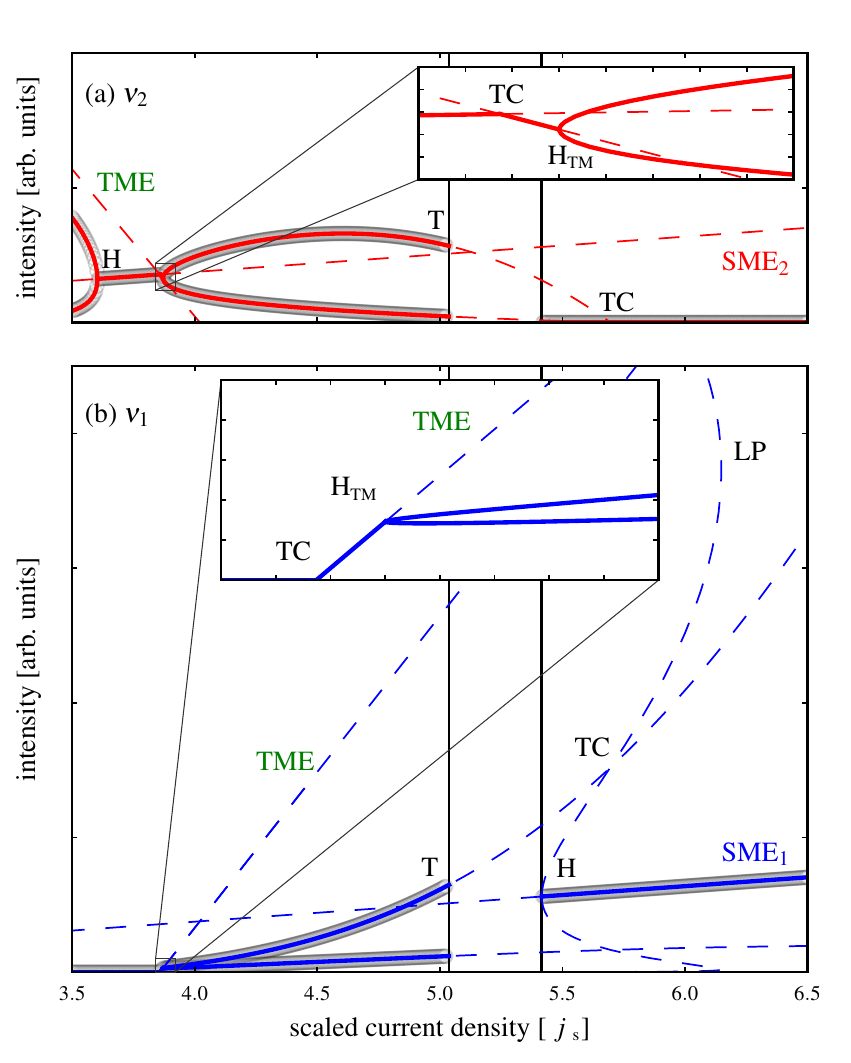}
  \caption{Bifurcation diagrams of $\nu_2$ (upper panel) and
    $\nu_1$ (lower panel). Single and dual-mode equilibrium states are labeled
    $\mbox{SME}_j$ and TME respectively. Bifurcation points are single-mode Hopf
    bifurcations (H), transcritical (TC), dual-mode Hopf bifurcations
    ($\mbox{H}_{\mbox{\tiny{TM}}}$), Torus bifurcations (T) and limit point
    (LP). Solid and dashed lines indicate stable and unstable objects
    respectively. The results of numerical simulations of the coupled system
    are included as a guide. Solid vertical lines bound the region of complex
    dual-mode dynamics.}
    \label{fig:tm_region_auto}
\end{figure}

To investigate this question and to shed more light on the bifurcation structure
in this system, in Fig. \ref{fig:tm_region_auto} we have plotted numerical
continuation results obtained using AUTO \cite{auto}. These results indicate
that the instability of the dual-mode limit cycle leading to complex dynamics is
due to a supercritical Torus bifurcation (T). We can also see that the dual-mode
limit cycle is present as an unstable object throughout the region of complex
dynamics, and that there is evidence for further bifurcations of interest
involving this object beyond the boundary of the region of complex dual-mode
dynamics. In particular, we can see that the unstable dual-mode limit cycle
collides with the unstable branch of single-mode SPs of mode $\nu_1$ in a
transcritical bifurcation (TC). This results in the unstable branch of
single-mode SPs becoming transversally stable for decreasing pump values until
the subcritical Hopf bifurcation of $\nu_1$ is reached. This narrow region of
transverse stability for this unstable branch of SPs may explain the sharp
feature near the relaxation oscillation frequency found in the power spectral
data of Fig. \ref{fig:exptgain}. The diminishing strength of this feature with
increasing current could be a result of the growth of the limit cycle away from
the location of the stable equilibrium.

Finally, we note that although a complete discussion of the bifurcation
structure that organises the dynamics of Fig. \ref{fig:sim_qs-qs} is beyond the
scope of the current paper, we can highlight a number of interesting parallels
with previous work on optically injected dual-mode devices \cite{osborne_09,
 blackbeard_12}. Mathematically, this system is also four dimensional, but it
features a single, three dimensional invariant manifold corresponding to the
single-mode injected system. Despite the lower symmetry of the dual-mode
injected system, in the example of \cite{osborne_09}, we also found a saw-tooth
structure characterized by symmetric but oppositely directed trajectories of the
two modes. These dynamics originated in a torus bifurcation of a dual-mode
periodic orbit. On the other hand, \cite{blackbeard_12} considered an example of
bursting dynamics from the region of the single-mode manifold. These dynamics
appeared near a cusp-pitchfork bifurcation of limit cycles and a curve of global
saddle-node heteroclinic bifurcations. We will present a similar two-parameter
bifurcation study of the current system and explore these connections further
in future work.



\section{Conclusions}
\label{sec:conclusions}
We have presented an experimental and theoretical study of the dynamics of a
dual-mode semiconductor laser with a saturable absorber. The device was a
specially engineered Fabry-Perot laser designed to support two primary modes
with a large frequency spacing. Fixing the voltage applied to the absorber
section, we performed a sweep in drive current in the gain section of the
device. We found that the dynamics evolved from familar self-pulsations in a
single mode of the device into a complex dynamical state of both modes. By
extending the well-known rate equation model for the semiconductor laser with a
saturable absorber to the multimode case, we were able to reproduce the observed
dynamics, and to show the fundamental role played by material dispersion in both
sections of the device in governing their appearance.

\textit{Acknowledgments.} The authors acknowledge the financial support of
Science Foundation Ireland under Grant ``SFI13/IF/I2785''.



\end{document}